\documentclass[fleqn,usenatbib]{mnras}

\usepackage{newtxtext,newtxmath}
\usepackage[T1]{fontenc}
\usepackage{ae,aecompl}
\usepackage{graphicx}	
\usepackage{amsmath}	
\usepackage{amssymb}	
\usepackage{booktabs}
\usepackage[table,xcdraw]{xcolor}
\usepackage{newtxtext,newtxmath}
\usepackage[T1]{fontenc}
\usepackage{ae,aecompl}
\usepackage{marvosym}
\usepackage{hyperref}
\hypersetup{colorlinks,allcolors=teal}
\usepackage{soul}
\usepackage{balance}

\newcommand{\beq}[1]{\begin{equation}\label{#1}}
\newcommand{\eeq}{\end{equation}}
\newcommand{\sub}[1]{_{\rm #1}}
\newcommand{\beqn}{\begin{equation}}
\newcommand{\eeqn}{\end{equation}}

\newcommand{\ap}{a\sub{p}}

\newcommand{\Mp}{M\sub{p}}

\newcommand{\Roo}{R\sub{o}}
\newcommand{\Rii}{R\sub{i}}
\newcommand{\Rp}{R\sub{p}}

\newcommand{\Ar}{A\sub{ring}}
\newcommand{\Ap}{A\sub{planet}}

\newcommand{\Mstar}{M\sub{\star}}

\newcommand{\nr}{\hat{n}\sub{r}}

\newcommand{\ns}{\hat{n}\sub{s}}
\newcommand{\tbright}{t\sub{bright}}
\newcommand{\tdark}{t\sub{dark}}

\definecolor{rev}{RGB}{50,109,255}

\usepackage{comment}

\title[Scattered light from ringed exoplanets]
{Scattered light may reveal the existence of ringed exoplanets
}
\author[Sucerquia et al.]{\parbox{\textwidth}{
Mario Sucerquia$^{1,2}$\thanks{E-mail:\href{mailto:mario.sucerquia@uv.cl}{mario.sucerquia@uv.cl}}, Jaime A. Alvarado-Montes$^{3,4}$, Jorge I. Zuluaga$^{5}$,\\ Mat\'ias Montesinos$^{1,2,6}$ and Amelia Bayo$^{2,1}$
}\vspace{0.4cm} \\
$^{1}$N\'ucleo Milenio Formaci\'on Planetaria - NPF, Universidad de Valpara\'iso, Av. Gran Breta\~na 1111, Valpara\'iso, Chile \\
$^{2}$Instituto de F\'isica y Astronom\'ia, Facultad de Ciencias, Universidad de Valpara\'iso, Av. Gran Breta\~na 1111, 5030 Casilla, Valpara\'iso, Chile\\
$^{3}$Department of Physics \& Astronomy, Macquarie University -- Sydney, NSW 2109, Australia.\\
$^{4}$Centre for Astronomy, Astrophysics, and Astrophotonics, Macquarie University -- Sydney, NSW 2109, Australia.\\
$^{5}$SEAP,  Instituto de F\'{\i}sica - FCEN, Universidad de Antioquia, Calle 70 No. 52-21, Medell\'in, Colombia.\\
$^{6}$Chinese Academy of Sciences South America Center for Astronomy, National Astronomical Observatories, CAS, Beijing 100012, China
}
\date{Accepted 2020 April 29. Received 2020 April 3; in original form 2020 January 25}
\pubyear{2020}

\begin{document}
\label{firstpage}
\pagerange{\pageref{firstpage}--\pageref{lastpage}}
\maketitle

\begin{abstract}

Rings around giant exoplanets (hereafter `exorings') are still a missing planetary phenomenon among the vast number of discovered planets. Despite the fact there exist a large number of methods for identifying and characterizing these exorings, none of them has been successful to date. Most of those efforts focus on the photometric signatures produced by rings around transiting exoplanets; thus, little interest has been intended for the detectable signatures that non-transiting ringed planets might cause owing to the excess of scattered starlight from both their atmosphere and the considerably large surface of their (hypothetical) rings-system. This extra scattering produced by exorings would occur at an orbital location defined here as the `the summer solstice' of a stellar light curve. In this letter, we develop a first-order model to estimate the photometric signatures of non-transiting exorings, and predict their detectability by using present and future facilities. We also show how, besides the discovery itself, our model can be used to constrain orbital and physical parameters of planet-rings systems.
\end{abstract}

\section{Introduction}

Thousands of exoplanets discovered so far are constantly unveiling new planetary phenomena and moving farther the limits of our knowledge. In not more than 35 years, humanity has discovered and characterized planets with masses and sizes spanning from a tenth of Mercury to tens that of Jupiter; and within a wide range of orbital configurations, compositions, and dynamical environments (see \citealt{Perryman2018}). All of these discoveries have been accomplished with the combined power of ground- and space-based telescopes and spectrographs (e.g. \textit{Kepler}, \textit{TESS}, \textit{SuperWASP}, \textit{HARPS}, etc); besides, the advent of valuable data from exoplanet projects/instruments such as \textit{PLATO}, \textit{CHEOPS}, \textit{SPHERE/ZIMPOL}, \textit{E-ELT}, \textit{GMT}, and \textit{JWST} (see e.g. \citealt{Lopez2019}) portends significant upcoming achievements.

In the last decade, the systematic monitoring of stellar light curves has proven high success for the detection and characterization of exoplanets. This technique, known as transit photometry, has enabled the study of planetary atmospheres (see e.g. \citealt{Batalha2017,Mayorga2019}),
detected exocomets and unusual objects (see e.g. \citealt{Rappaport2018}),
suggested the existence of yet-to-be-confirmed exomoons (\citealt{Teachey2018b}), found debris disks around sub-stellar objects \citep{kenworthy2015}; and may have the potential of revealing other previously unknown planetary processes as those in \citet{Boyajian2016} and \citet{Sucerquia2019}. Other methods like direct imaging and submillimeter interferometry have also confirmed their huge potential. In recent years, it has been announced the detection of complex circumstellar disks which hint at unusual planetary formation processes, as well as {\it circumplanetary discs} where moons or even planetary rings may possibly form (see e.g.
\citealt{Isella2019}).

Contrary to the outer planets of our Solar System, the detection of rings around giant exoplanets (`exorings' for short) remains elusive. These exorings may produce notorious photometric signatures if they orbit around transiting exoplanets (see \citealt{Zuluaga2015} and references therein), but no effort has fully succeeded at detecting the first exoring (see e.g. \citealt{Piro2019}).
However, transit photometric signatures are not the only viable method for identifying exorings, as it will be demonstrated later in this work.

In addition, by studying highly polarised scattered light from planetary surfaces, projects like \textit{RefPlanets} \citep{Hunziker2019} intend to perform the search, discovery, and characterization of evolved cold planets around nearby (bright) stars, making use of the instrument SPHERE/ZIMPOL attached to the VLT \citep{Schmid2018}. Scattered light has also been used for studying the transiting exoplanet HD 189733b \citep{Knutson2007}; and other authors have deeply studied similar methods to detect and fully characterize the atmosphere of direct-imaged exoplanets using polarized scattered light (see e.g. \citealt{Stolker2017}).

The stellar light curve of HD 189733 presented a primary and secondary transit, but a sinusoidal flux variation was also observed (see fig. 1 in \citealt{Knutson2007}). This variation was interpreted as a result of infrared emission from the day- and night-side of the planet and used to infer the temperature distribution across its surface. A similar study using visible wavelengths was conducted for one of the first transiting exoplanets studied by Kepler, HAT-P-7b \citep{Borucki2009}, and more recently \citet{Webber2015} have analysed the scattered light from Kepler-7b to gather information on its cloud-coverage and composition. Since then, several works (see e.g. \citealt{Williams2008},  \citealt{Zugger2010}, \citealt{Placek2013}, \citealt{Stolker2017} and references therein) have explored theories about the potential that detecting and measuring scattered starlight may have for discovering and characterizing exoplanets. 

Interestingly, \textit{none} of those works pondered over the idea that scattered light might also reveal the presence of planetary rings, which is precisely the purpose of this letter whose layout is as follows. In Section \ref{sec:refmod}, we develop a {\em toy model} of the scattered light from an exoring and its host planet. In Section \ref{sec:orb-par}, we describe how orbital and physical information of a non-transiting ringed planet can be obtained from its light curve. In Section \ref{sec:orbor}, we sketch a so-called {\em ring-centric model} of the light curve to be developed in a forthcoming paper. Finally, we conclude in Section \ref{sec:discussion} with a discussion of the model presented here and its possible applications.


\begin{figure}
	\includegraphics[width=\columnwidth]{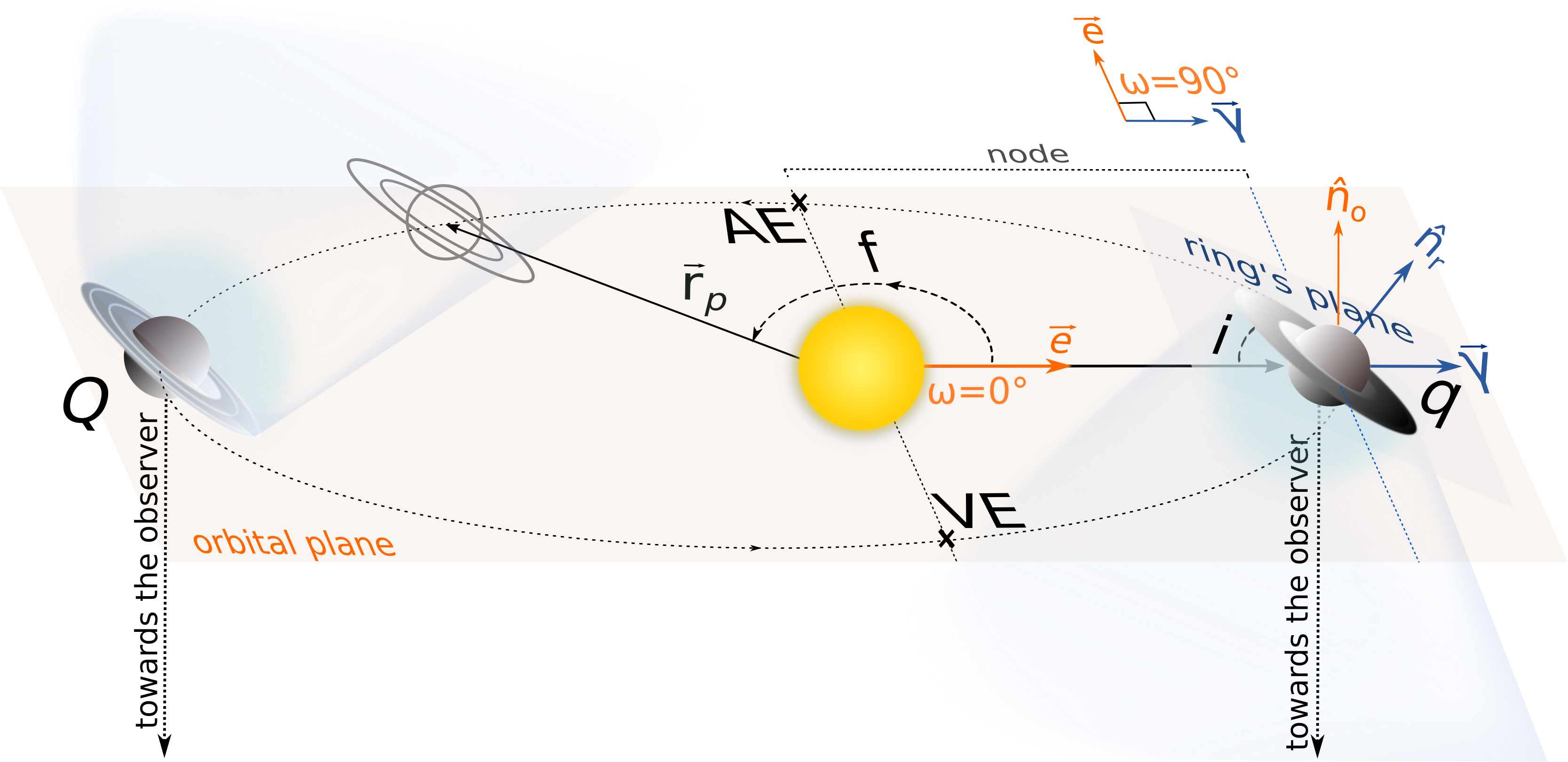}
    \caption{Schematic illustration of the system considered in our toy model. The shaded side of the ring is the one illuminated by the star. As the planet translates it crosses four critical points: the periastron ($q$ or summer solstice, when the maximum illuminated area of the ring is towards the observer); apoastron ($Q$ or winter solstice); and the Autumn (AE) and Vernal Equinoxes (VE), where both sides of the ring are dark. The direction of motion is counterclockwise with an orbital convention related to the Southern hemisphere.}
    \label{fig:scheme}
\end{figure}

\section{From orbital parameters to bright curve}
\label{sec:refmod}

Let us consider a planetary system composed of a ringed planet in a Keplerian orbit with semi-major axis $\ap$ and eccentricity $e$, orbiting around a main-sequence star with mass $\Mstar$ and luminosity $L_\star$ (see Fig. \ref{fig:scheme}). We include here only first-order `optical' effects, and assume a specific geometrical configuration wherein the planet's orbital plane is parallel to the sky plane (i.e. a face-on orbit); thus, the ring has a constant inclination $i$ to both planes. For instance, a value of $i=0$ means that the rings' plane coincides with the orbital and sky plane (i.e. the co-planar case).

Also, we define the \textit{orbital periastron argument}, $\omega$, as the angle between a vector $\vec{\gamma}$ which is normal to the line of the nodes resulting from the intersection between the rings' and orbital plane, and the Laplace--Runge--Lenz vector {$\vec{e}$} (whose magnitude is $e$). The line of the nodes is parallel to that one between the rings' Autumn (AE) and Vernal (VE) equinoxes which will be defined later. For the sake of simplicity, the rings' maximum area exposed to the star is assumed to happen at the passage by the periastron (point $q$ in Fig. \ref{fig:scheme}) and $\omega=0$°. In analogy with the Earth's seasons, we name this point the {\em summer solstice} of the visible ring surface. For a more detailed description of the geometry, the reader can refer to Section \ref{sec:orb-par} and Appendix \ref{ap:geometry}.

At first order, the amount of light released by the system at a determined time $I(t)$ is simply a superposition of direct starlight $I_\star$ and light diffusely scattered by the planet's atmosphere and rings,

\begin{equation}
I(t)=I_\star +\; [\alpha_\mathrm{p} \; \Ap + \alpha_\mathrm{r}\Phi(t)\, \Ar\cos i \sin i] \; B_\star(t),
\label{eq:ftot}
\end{equation}
where $B_\star(t)=I_\star/[4\pi r(t)^2]$ is the intensity at the instantaneous planet distance $r(t)=a[1-e \cos E(t)]$, with $E(t)$ the {\em eccentric anomaly} which is obtained by solving numerically the Kepler's equation of the planetary orbit; $\alpha_\mathrm{p,r}$ is the surface average {\em geometric albedo} of the planet (p) and ring (r); $\Ap$ and $\Ar\sin i \cos i$ are the isolated \textit{Lambertian} surface areas projected on the sky plane. The ring particles considered in our model are thick, porous, and irregular-shaped bigger than the considered wavelength, so no preferential direction of scattered light is expected.
However, small, dusty, and spherical particles with a size comparable to the wavelength must be treated according to Mie's solution \citep{Mie1908}.

The area of the ring is given by $\Ar=\mathrm{ \pi}  (\Roo^2 - \Rii^2)$, where $\Rii$ and $\Roo$ are the inner and outer radii, respectively. Furthermore, if we assume that the planet has a constant albedo, its Lambertian area will be simply: 

\begin{equation}
   \Ap= \int   \ns \cdot \, \mathrm{d}\vec s = \frac{\mathrm{\pi}}{2} \Rp^2, 
   \label{eq:aplan}
\end{equation}
with $\ns$ a unitary vector directed from a point on the planet's surface towards the star. The surface-integral in equation (\ref{eq:aplan}) is performed over the two illuminated octants of the planet, as seen from the observer's position (recall that a face-on orbit is assumed).

The quantity $\Phi(t)$ in equation (\ref{eq:ftot}) represents the variability of the cross-sectional ring area during a complete orbit, as seen from the star. At $q$ and $Q$ (see Fig. \ref{fig:scheme}) we have $\Phi=\Phi_\mathrm{max}$=1, whereas it is equal to zero at AE and VE. Quantitatively, $\Phi(t)$ is given by:

\begin{equation}
\Phi(t)=
\left\{
\begin{array}{cc}
\nr \cdot \hat{r}  & -\pi/2<f<\pi/2 \\
0 & \mathrm{otherwise} 
\end{array}
\right.
\label{eq:phi}
\end{equation}
where $\nr$ is the vector normal to the rings' plane, $\hat{r} = \Vec{r}/|\Vec{r}|$ the unitary vector pointing from the rings to the star, and $f$ the true anomaly of the planetary orbit.

Starting from equation (\ref{eq:ftot}) we can define the {\em normalised stellar flux anomaly}, $F(t)$, in our toy model as follows,

\begin{equation}
\label{eq:F}
F(t)\equiv\frac{I(t)-I_\star}{I_\star}=\frac{\alpha_\mathrm{r}\Phi(t)\, \Ar\sin i \cos i + \alpha_\mathrm{p} \, \Ap}{4\pi r(t)^2}
\end{equation}
which, as per usual, will be given in ppm (parts per million). However, strictly speaking, the planet's contribution to the light curve blends with that of the rings. This occurs because the planet's glow and shadow affect the scattered light from the rings (see Section \ref{sec:orbor}).

\begin{figure}
	\includegraphics[scale=0.33]{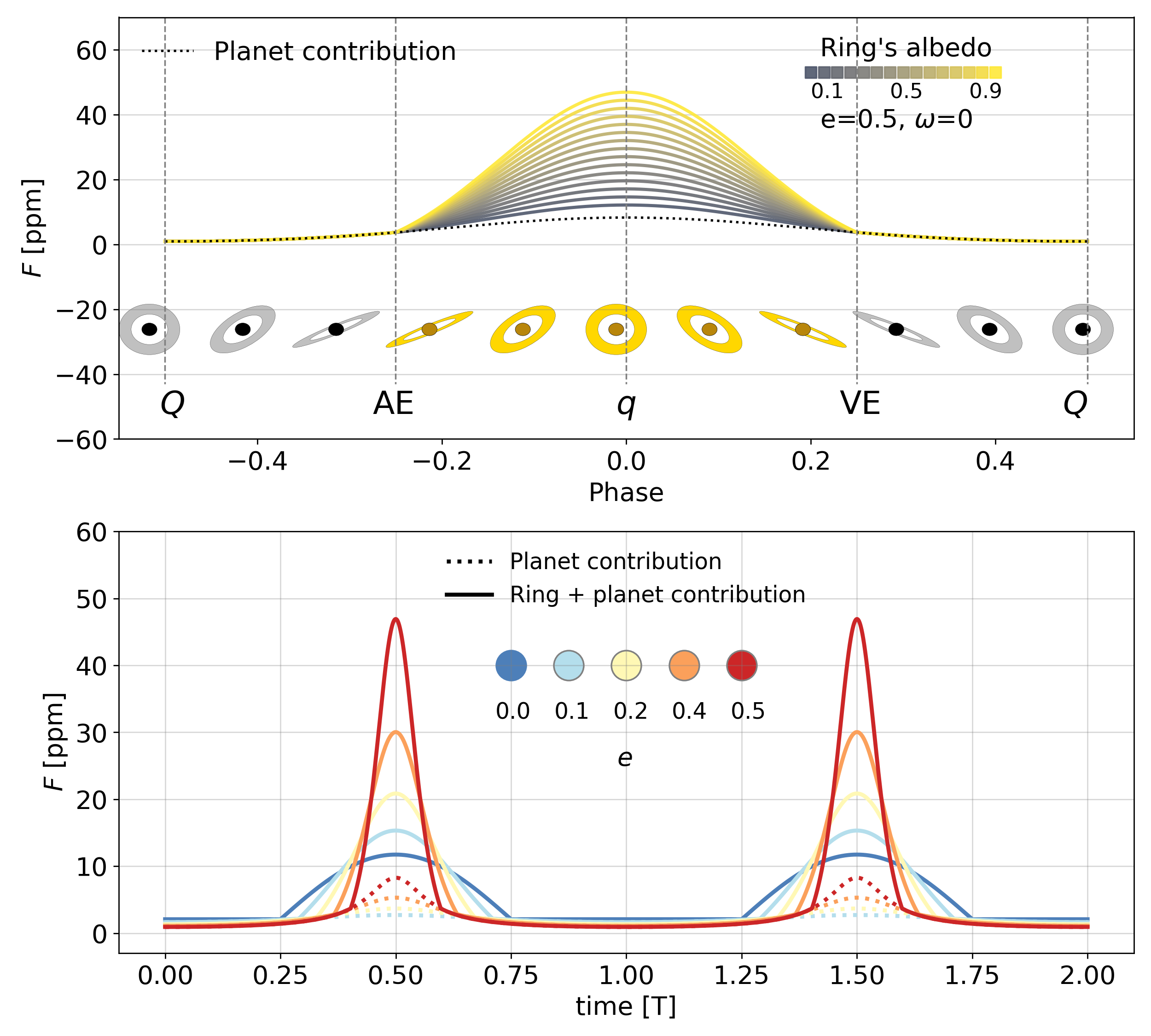}
    \caption{Upper panel: Normalised stellar flux of an exoring during one planetary orbit for albedos ranging from 0.1 to 1. Icons below the curve represent the cross-section of the rings-system as seen from the star. Colours of the icons indicate if the side towards the observer is illuminated (yellow) or not (gray). Lower panel: stellar light curve for two orbital periods showing the contribution of the planet (dotted lines) and planet+ring (solid lines).}
    \label{fig:fluxes}
\end{figure}

In the upper panel of Fig. \ref{fig:fluxes}, we show $F(t)$ for different orbital eccentricities in the case of a close-in Saturn, located at $\ap=0.1$ au from a Sun-like star; and assuming $\alpha_\mathrm{r,p}=1$ and $i=39^{\circ}$. It is worth mentioning that the actual axis tilt of Saturn's rings is $\sim 27$°, and $i\gtrsim39.2$° leads to Lidov-Kozai oscillations caused by stellar gravitational perturbations \citep{Sucerquia2017}.
The stellar light curve in the lower panel of Fig. \ref{fig:fluxes} allows us to identify four important properties: 1) the {\em maximum flux anomaly}, $F_\mathrm{max}$, or height of the peaks; 2) the {\em bright interval}, $\tbright$, or width of the peaks, corresponding to the time spent by the planet at VE and AE (when the illuminated side of the ring can be perceived by the observer); 3) the {\em dark interval}, $\tdark$, or time between AE and VE; and 4) the {\em dark flux anomaly}, $F_\mathrm{dark}$, defined either as the maximum or minimum value of $F$ during $\tdark$ (see Section \ref{sec:orb-par}). 

From the solid and dashed curves in the lower panel of Fig. \ref{fig:fluxes}, it can be noticed that the maximum flux anomaly of the ring is almost one order of magnitude larger than that of the planet, at least for the conditions considered in our toy-model (i.e. face-on orbit and summer solstice happening at periastron). This implies that for some current observing campaigns, the presence of rings around giant exoplanets could be detected when looking for scattered light, and at the same time, the absence of this conspicuous signal may allow constraining the occurrence rate of exorings among discovered extrasolar systems.

\begin{figure}
    \includegraphics[scale=0.33]{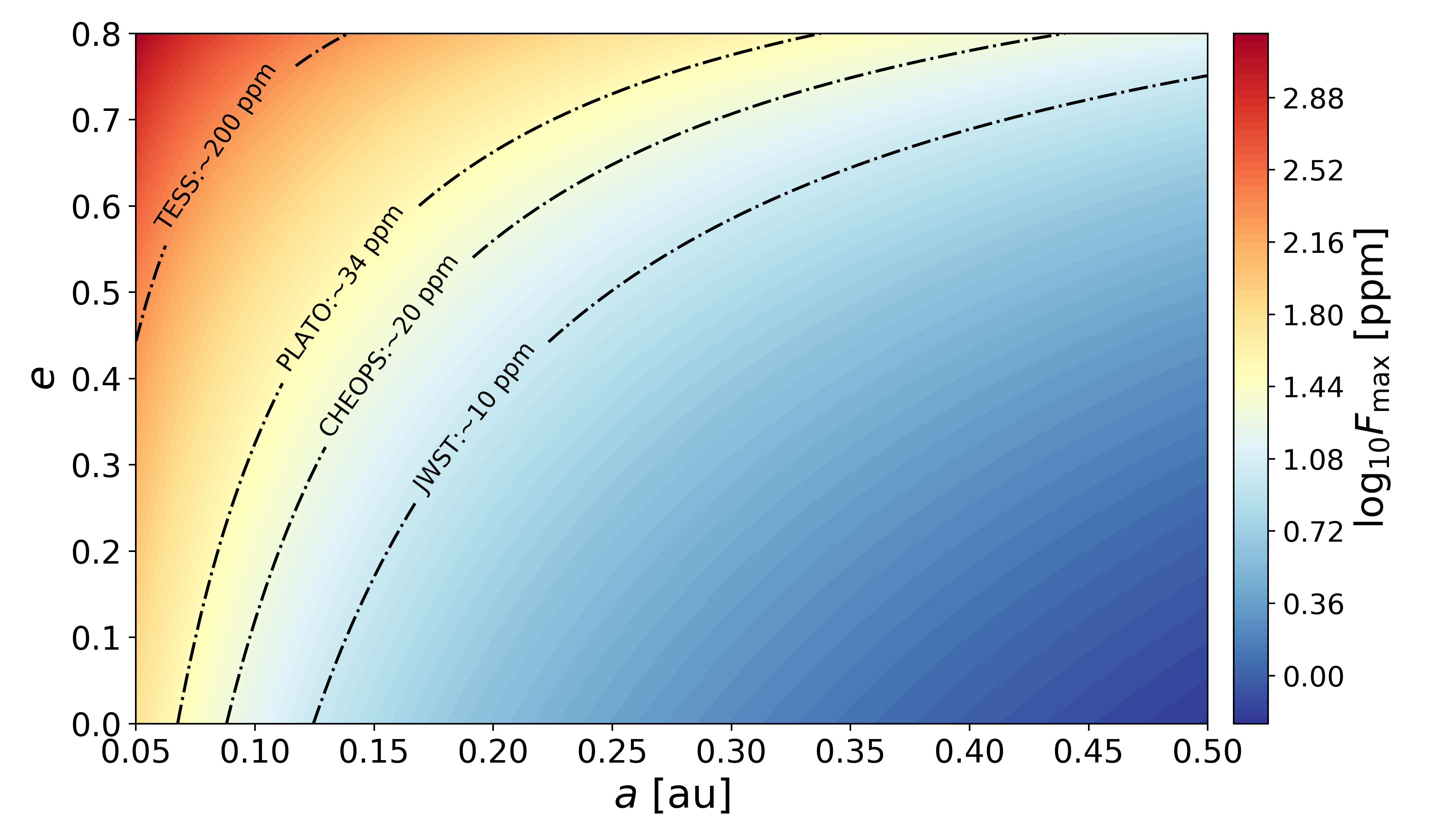}
    \caption{Contour plot of the normalised stellar flux anomaly $F$ in the orbital parameter space ($e-a$). We assume a Sun-like star and Kronian properties for the planet and its rings.}
    \label{fig:map_a_e}
\end{figure}

Contour plots of the maximum flux anomalies $F_\mathrm{max}$ are shown in Fig. \ref{fig:map_a_e} for a system with similar properties but different orbital configurations from those of Fig. \ref{fig:fluxes}. Within the conditions considered in the toy model presented here, Fig. \ref{fig:map_a_e} indicates that only those rings around close-in giant planets might be detected by \textit{TESS}, \textit{PLATO}, \textit{CHEOPS}, and \emph{JWST}. However, if future instruments achieve photometric sensitivities well below the $\sim10$ ppm level, cold planets (even beyond the snow-line) are theoretically detectable via scattered light from their system of rings.


\section{From bright curve to orbital parameters}
\label{sec:orb-par}

Can orbital and physical properties of non-transiting ringed planets be determined, estimated, or constrained from their light-curve attributes? Light curves certainly depend on many parameters in addition to those considered in our model; thus, the solution to this problem will be highly {\it degenerated}. However, we can still constrain the basic characteristics of the system through the four main properties defined in the previous section. At the research stage of this topic, where only the simplistic toy model of Section \ref{sec:refmod} has been developed, we will follow a parallel line of thought to the seminal work on transiting exoplanets by \citet{Seager2003}.

We assume that the mass of the star, $\Mstar$, is much larger than that of the planet, $\Mp$ (i.e. $\Mstar\gg\Mp$), and that $\Mstar$ can be estimated from independent methods such as the mass-luminosity relationship. Still, estimating accurate (higher than the mass-luminosity relationship) masses and radii for main-sequence stars from observations is not a straightforward task. It usually involves a combination of asteroseismology and/or interferometry, with observables calibrated against eclipsing binaries (see e.g. \citealt{Moya2018}). However, space-based missions are providing a very rich wealth of data that allow (or will allow shortly) for these precise estimates to be calculated in an almost routine fashion.

The planet's orbital period, $T$, will be equivalent to the interval between light-curve peaks (see Fig. \ref{fig:fluxes}), so the semi-major axis is computed as $a \approx [G \Mstar (T/2\pi)^2]^{1/3}$ (i.e. Kepler's third law). In our toy model and under the assumption of a pure Keplerian undisturbed orbit, the width of the bright peaks in the light curve is the time from VE ($f=-\pi/2$) to AE ($f=\pi/2$) and is simply given by:
\begin{equation}
\label{eq:time-e-inverted}
\frac{t_\mathrm{bright}}{T} =\hspace{-0.5pt} \frac{1}{2\pi}\left[\cos^{-1} \eta_\mathrm{+} + \cos^{-1} \eta_\mathrm{-} - e\left(\sqrt{1-\eta_\mathrm{+}^2}+\sqrt{1-\eta_\mathrm{-}^2}\right)\right]    
\end{equation}
where
\begin{equation}
\label{eq:eta}
\eta_{\pm} = \frac{e\pm\sin \omega}{1\pm e\sin \omega},
\end{equation}
with $\omega$ the periastron angle. If $\omega=0$°,
\begin{equation}
\label{eq:time-e-trascendental}
\frac{\tbright}{T}=\frac{1}{\mathrm{\pi}} \left[\cos^{-1}{(e)}-e\sqrt{1-e^2}\right], 
\end{equation}
which can be numerically inverted to provide $e$ as a function of $\tbright$. For small values of $e$, we can write an analytical expression for $e$ as a function of $\tbright$ (see the dashed line in Fig. \ref{fig:time-e}):

\begin{figure}
	\includegraphics[width=\columnwidth]{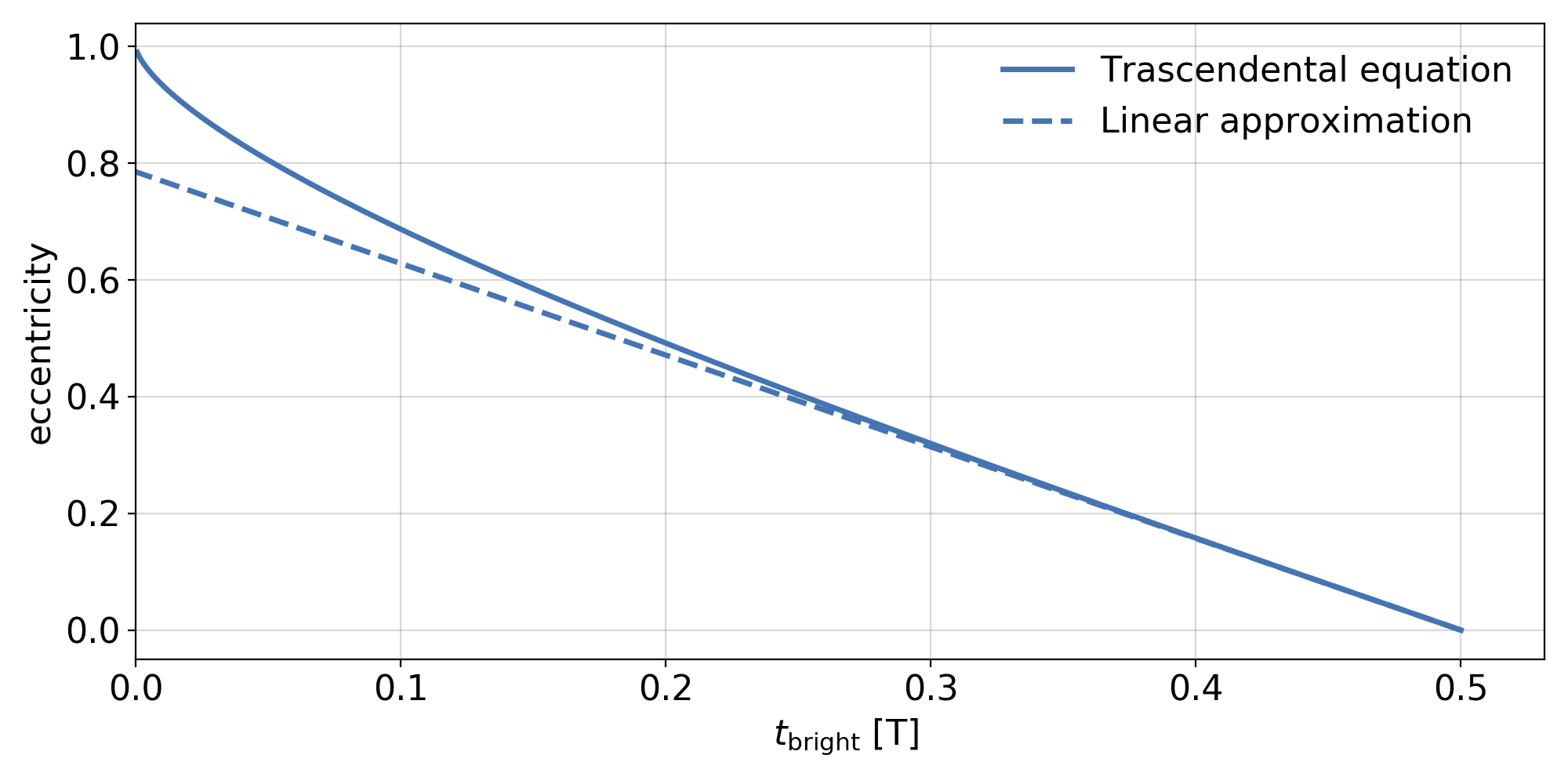}
    \caption{Planetary eccentricity ($e$) as a function of the bright time ($\tbright$) in the toy model developed in this work (see equation \ref{eq:time-e-trascendental}).
    \label{fig:time-e}}
\end{figure}
\begin{equation}
\label{eq:elinear}
e\approx\frac{\pi}{2}\left(\frac{1}{2}-\frac{\tbright}{T}\right).
\end{equation}
This linear approximation always underestimates the actual value of $e$, and up to $e\approx 0.5$ it has relative errors below $5$ per cent.

To estimate the planet's size (or more precisely its Lambertian area) we can use the dark flux anomaly $F_\mathrm{dark}$. From equation (\ref{eq:ftot}), between AE and VE, the flux $F(t)$ is constrained by:
\begin{equation}
\label{eq:Fdark}
F(t)\leq F_\mathrm{dark}\equiv\frac{\alpha_\mathrm{p} \Ap}{4\pi a^2(1-e^2)^2}.
\end{equation}
Since $\alpha_\mathrm{p}\leq 1$ we can constrain the planetary Lambertian area:
\begin{equation}
\label{eq:Apmin}
\frac{\Ap}{a^2}\geq \frac{A_\mathrm{planet,min}}{a^2}\equiv 4\pi\,(1-e^2)^2F_\mathrm{dark}.
\end{equation}
In our toy-model (see Section \ref{sec:refmod}), $F_\mathrm{max}$ occurs at periastron $q=a \, (1-e)$. Thus, according to equation (\ref{eq:ftot}) $F_\mathrm{max}$ is given by
\begin{equation}
\label{eq:Fmax}
F_\mathrm{max} = \frac{\Delta_\mathrm{pr}}{4\pi\,(1-e)^2},
\end{equation}
where,
\begin{equation}
\label{eq:deltapr}
\Delta_\mathrm{pr} = \alpha_\mathrm{r} \frac{\Ar}{a^2} \cos i \sin  i + \alpha_\mathrm{p} \frac{\Ap}{a^2}.
\end{equation}
As $\alpha_\mathrm{r},\alpha_\mathrm{p}\leq 1$, $\cos i \sin  i <1$, and $\Ap<A_\mathrm{planet,min}$ we have,

\begin{equation}
\label{eq:Armin}
\frac{\Ar}{a^2} \geq \frac{A_\mathrm{ring,min}}{a^2} \equiv 4\pi (1-e^2) F_\mathrm{max}-\frac{A_\mathrm{planet,min}}{a^2}
\end{equation}
To sum up, if we measure $F_\mathrm{max}$, $F_\mathrm{dark}$, and $\tbright$ from the light curve of a non-transiting ringed exoplanet, we can compute $e$ (equation \ref{eq:elinear}), $A_\mathrm{planet,min}/a^2$ (equation \ref{eq:Apmin}), and $A_\mathrm{ring,min}/a^2$ (equation \ref{eq:Armin}). Additionally, if we find $a$ from $T$ and $\Mstar$, an estimation of the planet and rings' radii can also be obtained.


\begin{figure}
	\includegraphics[scale=0.32]{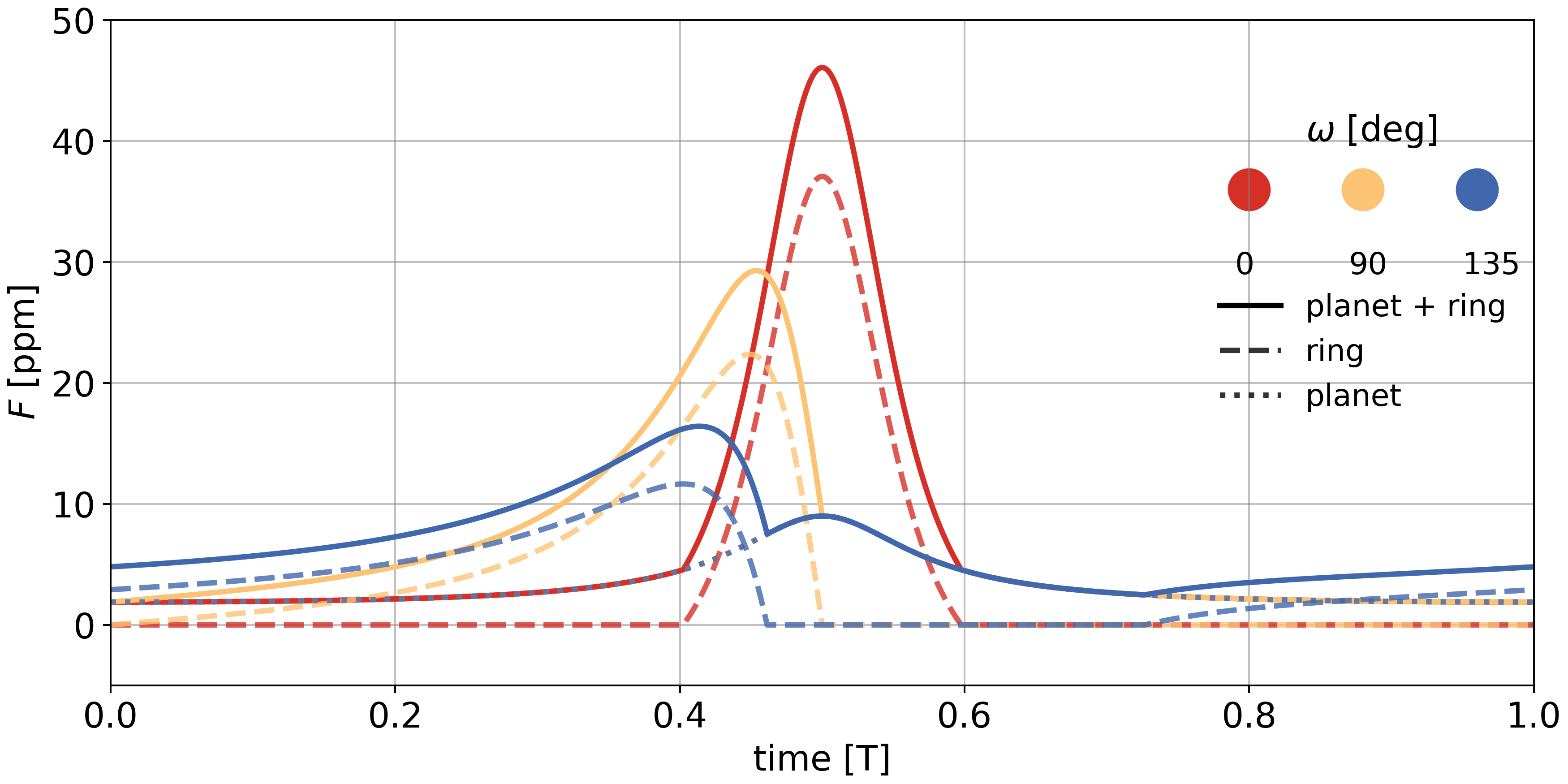}\vspace{0.1cm}
	\includegraphics[scale=0.32]{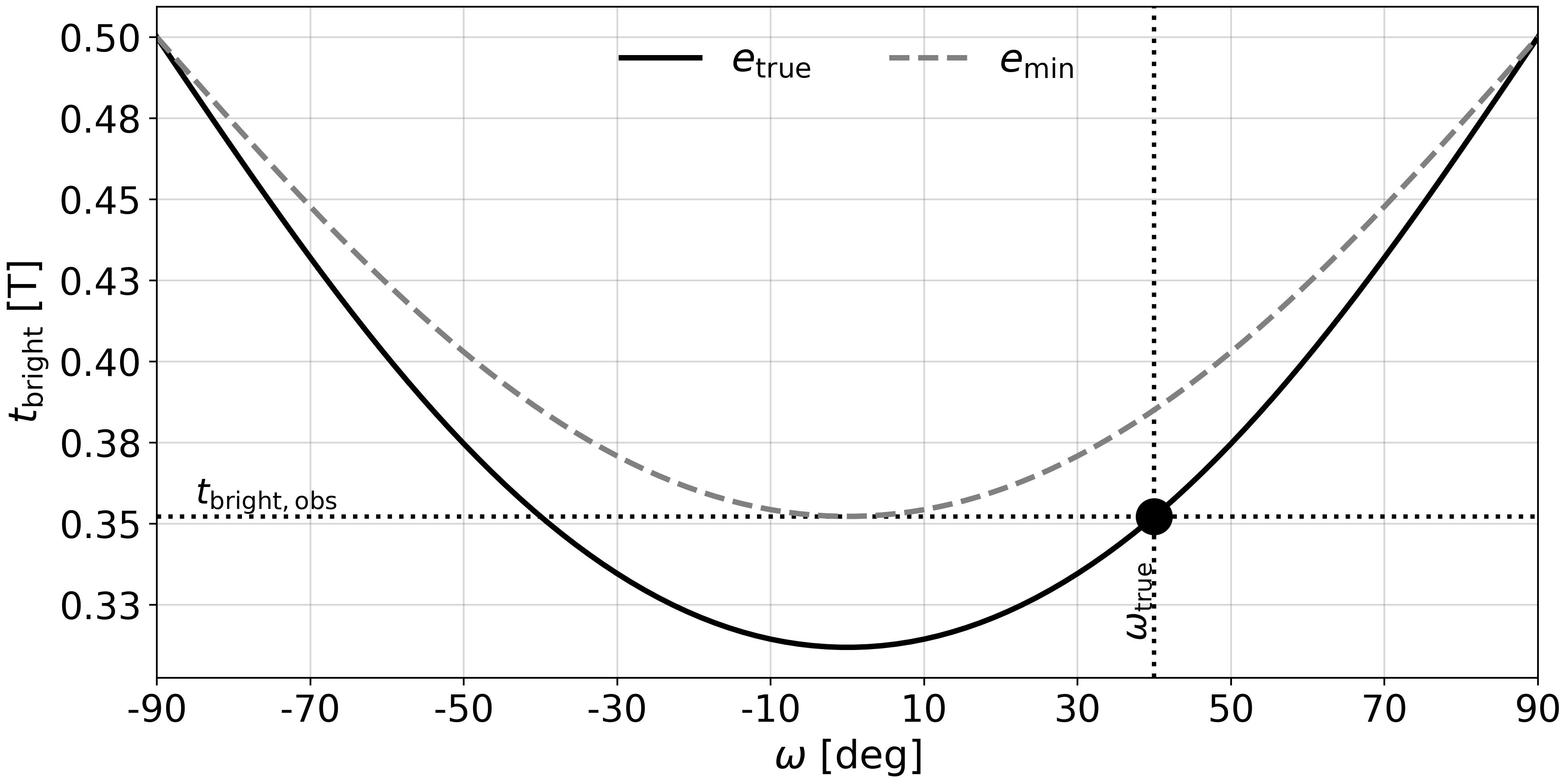}
    \caption{Upper panel: Normalised flux as a function of time. The contribution of the planet (dotted lines), ring (dashed lines), and planet+ring (solid lines) are shown. The colours represent different values of the orbital periastron argument ($\omega$) in Fig. \ref{fig:scheme}. Lower panel: relationship between the bright time ($\tbright$) and the estimated eccentricity ($e$) given by equation (\ref{eq:time-e-inverted}).}
    \label{fig:wsys}
\end{figure}
\section{Beyond the toy model}
\label{sec:orbor}

The toy model developed so far has assumed a very specific orientation for the orbit of a non-transiting planet and its rings. This simplifies significantly the analytical description of its light curve in a similar avenue to the original modelling of transit light curves by \citealt{Seager2003}. Two basic quantities, however, will certainly be different in real systems: 1) $\omega$ respect to the summer solstice (i.e. point q in Figs \ref{fig:scheme} and \ref{fig:fluxes}); and 2) a non-zero inclination $i$ of the orbit respect to the sky plane.

The effect that $\omega$ has on the light curve is shown in the upper panel of Fig. \ref{fig:wsys}. It is evident that as the summer solstice shifts its position from the periastron, the amplitude of the signal decreases and reaches a minimum when the summer solstice occurs at $\omega=180^\circ$ (i.e. the apoastron). Still, the light curve preserves its main features having a fainter $F_\mathrm{max}$, a larger $F_\mathrm{dark}$, and a $t_\mathrm{bright}$ much harder to measure, as compared to the toy model. 

In the lower panel of Fig. \ref{fig:wsys} we analyse how $\omega$ affects the relationship between $\tbright$ and $e$ (equation \ref{eq:time-e-inverted}). As expected, $\tbright$ is minimum for the toy model ($\omega=0^\circ$) and reaches a maximum value when the summer solstice occurs at quadrature ($\omega=90^\circ, -90^\circ$). Since an optimal orbital configuration was assumed in the toy model, equation (\ref{eq:elinear}) and the observed $\tbright$ will underestimate the true eccentricity of the orbit (see the dashed curve in the lower panel of Fig. \ref{fig:wsys}). However, even in a more realistic situation, equation (\ref{eq:elinear}) is still valid as a formula for constraining $e$:

\begin{equation}
\label{eq:emin}
e\geq e_\mathrm{min}=\frac{\pi}{2}\left(\frac{1}{2}-\frac{\tbright}{T}\right), 
\end{equation}
Also, an unknown orbital inclination contributes either to increasing or reducing $F_\mathrm{dark}$ and $F_\mathrm{max}$ respect to their corresponding values in the toy model (equations \ref{eq:Fdark} and \ref{eq:Fmax}, respectively). In the case of $F_\mathrm{dark}$, $\omega$ tends to increase its true value, while $i$ may either reduce it or increase it,  compensating thereby the effect of $\omega$. On the other hand, the constraint given by equation (\ref{eq:Apmin}) will not be substantially modified (at least within a first-order estimation of $\Ap$). By comparison with the toy model both $\omega$ and $i$ lower the measured value of $F_\mathrm{max}$. This implies that equation (\ref{eq:Armin}) would remain valid, although the minimum $\Ar$ will probably be underestimated.

To find {\em fiducial} parameters of real systems, a more general and complex model for fitting stellar light curves needs to be developed. To that end, we should include second-order effects such as the shadow/shine of the planet's surface and atmosphere; forward scattering; the opacity of the rings; the polarization and reflection of light by dusty/icy ring-particles, etc. For this {\em optical complexity} we propose (and intend) to create a scattering model centred on the ring (most models are centred on the star), where the properties of the scattered light are computed in the sky above the planetary rings after describing the astronomical position of the star, the planet, and the observer. However, such a model is ongoing research that will be introduced in a forthcoming paper.


\section{Summary and conclusions}
\label{sec:discussion}

The first-order semi-analytical model presented here demonstrates that rings may produce detectable and distinguishable photometric signatures on the light curve of non-transiting exoplanets. We have quantified the phase-dependent amount of light that a tilted ringed planet can reflect towards a distant observer, and the model is described through four main morphological features of the so-called \textit{bright curve} that help us derive some analytical constraints to orbital and physical parameters of the system. This model can be useful to detect (ringed) planets with large ring surfaces lying in face-on orbits, whose survival time-scale is $\sim1$ Myr \citep{Schlichting2011}. Thus, our method is complementary to transit photometry which is rather biased in favour of edge-on configurations.

It is worth noting that we considered a system composed of a Saturn-like planet in a close-in orbit around a Sun-like star. This is a simplified scenario for the sake of presenting our model, but under this assumption, transits are still likely to occur: for an orbit of 0.1 au, a Saturn-like planet would have a transit probability of $\sim10$ per cent. However, the remaining 90 per cent of the exoplanets of this specific population cannot be detected via transit photometry, so the model proposed here would be ideal for complementing the general distribution of close-in exoplanets.

The applicability of our model would be further ensured if we had independent \textit{astrometric} measurements of stellar masses ($\Mstar$). However, astrometry is still a developing area with five candidates awaiting confirmation and only ten confirmed planets. By following up on those systems and using the inferential procedure presented in Section \ref{sec:orb-par}, one might reveal the presence of ringed planets and help constrain some of their characteristics. Moreover, the next releases of data from GAIA will increase the number of discoveries via astrometry and thus the chance of finding and characterizing the yet-unseen population of exorings.

Within the scope of this work, and after some assumptions on the radius of the central body and the shape/orientation of the rings, we simulated the system 1SWASP J140747 \citep{kenworthy2015}. Regardless of its low stellar luminosity and large planetary semi-major axis ($\ap \sim 3.9$ au), we found that the maximum amount of scattered light from the system rounds the 150-ppm level at the orbital periastron ($e \, \sim 0.6$), becoming detectable with current instrumentation. Also, in analogy with HD 189733b \citep {Knutson2007}, a modulation between consecutive stellar flux-peaks should be expected. This is produced due to the scattered light from the ring which induces a phase-modulated shift in the stellar light curve.

The signal produced by scattered light from face-on close-in planets may be meagre and misleading when the signal-to-noise ratio (SNR) of observations is considerably low. This, however, can be circumvented by adopting state-of-the-art methods to increase the SNR (e.g. \citealt{Colon2012, Stefansson2017}). With high instrumental sensitivity levels, phase-dependent deviations of the scattered starlight only from the planet (see the dotted lines in the lower panel of Fig. \ref{fig:wsys}) can help constrain different orbital parameters using analytical expressions of the variable flux, $F(t)$.

A striking result to the phenomena described in this work is that a bright curve resulting from the scattered light from planetary rings is an indirect way of characterizing the planet's interior/atmosphere, which is apposite in the \textit{TESS} (present) and \textit{JWST} (future) observational era. Because of its broad wavelength coverage and high sensitivity, the \textit{JWST} will be a powerful tool for exoplanet transit spectroscopy that should allow us to achieve the required precision of 10–100 ppm either for transit or flux-peaks detection.

Finally, if we measure the polarized light of aligned dust-grains in the rings, embedded into an undetected exoplanet's magnetic field, we can also constrain the planet's rotation and composition. The evidence of those magnetic fields may help verify and improve current models of magnetic field formation and evolution. To summarize, the still unknown population of exorings, in particular, those discovered by scattered light, might be a proxy whereby we can further characterize and constrain the orbital and physical properties of close-in giant exoplanets.

\section*{Acknowledgements}
The authors thank the referee, Dr. Sebastien Charnoz, whose insightful comments helped improve this work. MS, AB, and MM are supported from Iniciativa Cient\'ifica Milenio (`ICM') via N\'ucleo Milenio de Formaci\'on Planetaria. JAA-M is funded by the International Macquarie University Research Excellence Scholarship (`iMQRES'), and JIZ by Vicerrector\'ia de Docencia, UdeA. MM is supported by the Chinese Academy of Sciences through a CAS-CONICYT Postdoctoral Fellowship administered by the CAS South America Center for Astronomy (CASSACA), Chile. AB thanks support from FONDECYT grant 1190748. Thanks to Johan Olofsson and the NPF team for helpful discussions.




\bibliographystyle{mnras}
\bibliography{references.bib} 


\appendix

%


\vspace{-1cm}
\section{System Geometry}
\label{ap:geometry}

\begin{figure*}
	\includegraphics[width=16 cm]{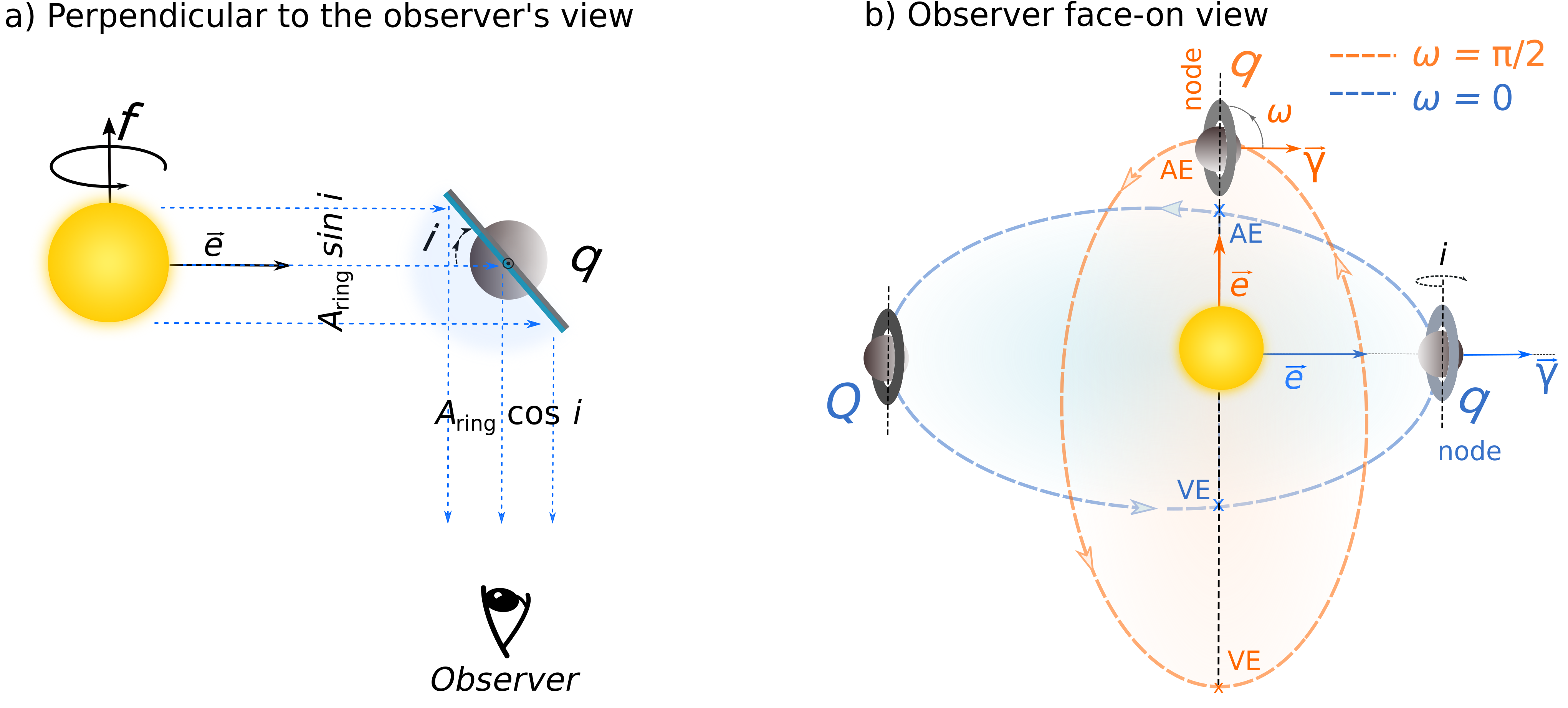}
    \caption{a) Schematic illustration of the system considered in this work. b) Schematic of two different orbital orientation on the sky-plane.    \label{fig:geometry}}
\end{figure*}

For study purposes, the system considered in this letter is composed of a nearby ringed planet whose orbital plane is co-planar to that of the sky. The rings-system around the planet has a constant inclination $i$, as it is shown in Figs \ref{fig:scheme} and \ref{fig:geometry}, where $i=0$ means that the rings' and orbital plane are co-planar. Also, the system studied in Section \ref{sec:orbor} assumes an orbital orientation $\omega=0$° where the Laplace--Runge--Lenz vector $\hat{e}$, whose magnitude is the orbital eccentricity, is aligned with $\vec{\gamma}$ which is a vector perpendicular to the line of the nodes produced by the intersection between the orbital and ring's plane (see the blue orbit in the left-hand panel of Fig. \ref{fig:geometry}). Also, an orange orbit is depicted in the right-hand panel of Fig. \ref{fig:geometry}, but in this case $\omega=90$°, meaning that $\vec{\gamma} \perp \vec{e}$.

An equivalent approach consists in to rotate the ring around a vector normal to the plane which passes through its centre. In any case, $\omega$ pinpoints the changes of the ring's orientation leaving the inclination $i$ constant. However, we choose the former definition (i.e. that of the previous paragraph) since it allows us to derive/constrain other orbital parameters such as the orbital eccentricity by using information that comes out from the reflected light by the ring.

Regarding the above, $\omega=0$° (where the planetary true anomaly $f=0$) makes the rings to appear brighter at the periastron as compared to other configurations, since the area of the rings projected towards the star is maximum ($A_\mathrm{ring} \sin{i}$). In other points of the orbit (i.e. $f \neq 0$) such a projected area is modulated by $\Phi$, defined in equation (\ref{eq:phi}), which decreases during the equinoxes where $\hat{n}_\mathrm{s} \cdot \hat{r} = \cos{f}= \cos{\pm \pi/2}=0$. Thus, the time-dependent projected area of the rings towards the star is defined as $A_\mathrm{ring}(t) = A \,\sin i\, \Phi(t)$. It is important to note that the line connecting the equinoxes passing through the star (see the black dashed-line in Fig. \ref{fig:scheme}) is parallel to the line of the nodes defined previously.

To conclude, since the system is assumed to be face-on aligned, only a portion of the reflected photons reaches the observer. To account for this effect we need to multiply the aforementioned projected area by $\cos{i}$, so we obtain that the intensity of the observed light, presented in equation (\ref{eq:F}), must be given by:
\balance

\begin{equation}
I_\mathrm{ring}(t) = I_\star A_\mathrm{ring}(t)B_\star(t) =  \frac{I_\star}{4 \, \pi r(t)^2} \, A \,\sin {i} \cos{i} \, \Phi(t)    
\end{equation}

\bsp	
\label{lastpage}
\end{document}